\documentclass[12pt,preprint]{aastex}
\usepackage{xspace}
\received{}
\revised{}
\accepted{}

\newcommand\msun{\ensuremath{M_\sun}\xspace}
\newcommand\mcrit{\ensuremath{M_{\rm crit}}\xspace}
\newcommand\teff{\ensuremath{T_{\rm eff}}\xspace}
\newcommand\logg{\ensuremath{\log g}\xspace}
\newcommand\ubv{\ensuremath{U\!BV}\xspace}

\slugcomment{Accepted for publication in the Astrophysical Journal Letters} 
\shorttitle{Massive white dwarfs in NGC 2168}
\shortauthors{Williams, Liebert \& Koester}

\begin{document}

\title{An empirical initial-final mass relation from hot,
  massive white dwarfs in NGC 2168 (M35)}  

\author{Kurtis A. Williams}
\affil{Steward Observatory}
\affil{933 N. Cherry Ave., Tucson, AZ 85721}
\email{kurtis@as.arizona.edu}
\author{M. Bolte}
\affil{UCO/Lick Observatory}
\affil{University of California, Santa Cruz, CA 95064}
\email{bolte@ucolick.org}
\and
\author{Detlev Koester}
\affil{Institut f\"ur Theoretische Physik und Astrophysik}
\affil{Universit\"at Kiel, 24098 Kiel, Germany}
\email{koester@astrophysik.uni-kiel.de}
\begin{abstract}
The relation between the zero-age main sequence mass of a star and its
white-dwarf remnant (the initial-final mass relation) is a powerful
tool for exploration of mass loss processes during stellar evolution.
We present an empirical derivation of the initial-final mass relation
based on spectroscopic analysis of seven massive white dwarfs in NGC
2168 (M35).  Using an internally consistent data set, we show that the
resultant white dwarf mass increases monotonically with progenitor mass for
masses greater than 4\msun, one of the first open clusters to show this
trend.  We also find two massive white dwarfs foreground to the
cluster that are otherwise consistent with cluster membership.  These
white dwarfs can be explained as former cluster members moving
steadily away from the cluster at speeds of $\lesssim 0.5$ km/s since
their formation and may provide the first direct evidence of the
loss of white dwarfs from open
clusters. Based on these data alone, we constrain the upper mass
limit of WD progenitors to be $\geq 5.8\msun$ at the 90\% confidence level
for a cluster age of 150 Myr. 
\end{abstract}

\section{Introduction}
White dwarfs (WDs) are the final state of stellar evolution for
the vast majority of intermediate- and low-mass stars.  The upper mass
limit of WD progenitor stars, \mcrit, is also the lower mass
limit of core-collapse supernova progenitors.  The current best observational
estimate for \mcrit comes from spectroscopic analysis of WDs in the open cluster
\objectname{NGC 2516} \citep{Koester1996} and is 
$\mcrit\approx 8\msun \pm 2\msun$.  Due to the steepness of the
initial-mass function (IMF), however, this range results in a factor
of $\approx 2$ uncertainty in the number of supernovae and duration of
supernova-driven winds resulting from bursts of star formation. This
uncertainty in turn has a large impact on understanding the
star-formation rate in galaxies 
 \citep[e.g.][]{Somerville1999}, the evolution of
starbursts \citep[e.g.][]{Leitherer1999}, and  the fate
of low-mass dwarf galaxies at early times \citep[e.g.][]{Dekel1986}.

The best observational constraints on \mcrit are obtained from studies of
WD populations in open clusters with ages $\lesssim 150$ Myr
\citep{Williams2002}.  \objectname{NGC 2168} (M35) is one of the
richest, compact and nearby open clusters in this age range, with age
determinations ranging from $\sim 100$ Myr \citep{vonhippel2000} to
$\sim 200$ Myr \citep{Sung1999}.  The WD cooling sequence of
NGC 2168 has been discussed often in the literature, with recent photometric
analyses by \citet{vonHippel2002} and \citet{Kalirai2003}.  

With spectroscopy of cluster WD candidates, it is possible to
determine unambiguously if the objects are WDs and, for the bona fide WDs, to 
determine \teff and \logg, and to derive
the cooling age ($\tau_{\mathrm cool}$), mass and luminosity.  
For those WDs with cooling ages smaller than the cluster age and
distance modulus consistent with cluster membership, subtraction 
of the WD cooling age from the age of the
open cluster results in the lifetime of the progenitor star, and
stellar evolutionary models can then be used to determine the
progenitor star mass.  \citet{Reimers1988} applied this
algorithm to two NGC 2168 WDs candidates identified on photographic plates by
\citet{Romanishin1980}; however, their signal-to-noise was too low to
determine the WD properties precisely.

As part of our ongoing program to identify and spectroscopically
analyze the WD populations of open clusters, the Lick-Arizona White
Dwarf Survey \citep[LAWDS,][]{Williams2004}, we
have obtained high signal-to-noise spectra of eight candidate
massive WDs in NGC 2168.  Six or seven of the observed WDs are hot, high-mass
WDs likely to be cluster members.  This sample doubles the number of
known open cluster WDs with high-mass progenitors.
A detailed photometric and spectroscopic 
analysis of the entire WD population of this cluster will be presented
in a later paper. In this Letter we present the constraints on \mcrit and
the upper end of the initial-final mass relation for WDs based on the 
high-mass WDs already analyzed.

\section{Observations \& Analysis}
\ubv imaging of NGC 2168 was obtained in September 2002 with
the Lick Observatory Shane 3-m reflector and the PFCam prime-focus
imager; additional \ubv imaging of a larger field center on the
cluster was obtained in
January 2004 with the KPNO 4m MOSAIC camera.  PSF-fitting photometry
was obtained using the DAOPHOT II program \citep{Stetson1987}.
Candidate WDs were selected by their blue excess in \ubv color space.
Figure~\ref{fig.cmd} shows the
color-magnitude diagram of all objects detected in $U$, $B$, and $V$
across the entire MOSAIC field.
Several very blue, faint objects are observed in the diagram; these 
are our candidate WDs. All four WD candidates of
\citet{Romanishin1980} are recovered, as 
is the WD candidate in \citet{vonHippel2002}.  Astrometry and
photometry for the WD candidates in this Letter are presented in
Table~\ref{tab.phot}. 

Spectroscopic observations of selected WD candidates were obtained
with the blue camera of the LRIS spectrograph on the Keck I
10-m telescope \citep{Oke1995}. A 1\arcsec-wide longslit at
parallactic angle was used with the 400 l mm$^{-1}$, 3400\AA~blaze
grism for a resulting spectroscopic resolution of $\sim 6$\AA.  
The spectra were extracted and a relative spectrophotometric
calibration applied using standard \emph{IRAF}
\footnote{\emph{IRAF} is
  distributed by the National Optical Astronomy Observatories, which
  are operated by the Association of Universities for Research in
  Astronomy, Inc., under cooperative agreement with the National
  Science Foundation} 
routines.  

The \teff and \logg were determined for each WD using simultaneous
Balmer-line fitting \citep{Bergeron1992}.  The model spectra are
updated versions of those in \citet{Finley1997}.  The WD
evolutionary models of \citet{Wood1995} were used to calculate the
mass ($M_{\rm WD}$) and cooling age ($\tau_{\rm WD}$) of each WD.  A distance
modulus to each WD was measured by comparing the observed $V$-magnitude 
to the absolute magnitude $M_V$ calculated from the
best-fitting model atmosphere and the appropriate WD cooling model.
Errors in the fits were determined empirically
by adding the noise measured for each spectrum to the best-fitting
model spectrum convolved with the instrumental response. These
simulated spectra were fit by the same method; 
nine iterations were used to calculate  the
scatter in \teff, \logg, $M_{\rm WD}$ and $\tau_{\rm WD}$.  The fitting
procedure is discussed in depth in an upcoming paper on the open
clusters \objectname{NGC 6633} and \objectname{NGC
  7063} \citep{Williams2004}.  The atmospheric 
fits and derived WD masses and ages are given in
Table~\ref{tab.spec}. The Balmer line fits are shown in
Figure~\ref{fig.linefits}.  

A systematic error in the fits of the
hottest ($\teff\gtrsim 50000{\rm K}$) WDs became apparent, as it was
not possible to simultaneously fit all the Balmer lines.  In these
cases, the fits were limited to H$\beta$, H$\gamma$, and
H$\delta$. For LAWDS 22, no satisfactory convergence was achieved with
these limited fits; the best-fit models are used in the
analysis. WDs with $\teff\gtrsim 50000$K are known to exhibit metals
in the atmosphere and NLTE effects \citep[e.g.][]{Napiwotzki1992,Holberg1998}, 
neither of which are included in
the models.

The progenitor mass for each WD was calculated by subtracting the WD
cooling age from the cluster age.  The age difference is the total
lifetime of the progenitor star.  The lifetimes of stars as a
function of mass and metallicity are calculated from the stellar
isochrones of \citet{Girardi2002}.  The progenitor mass for each
WD likely to be a cluster member
is given in Table~\ref{tab.masses} for an assumed cluster age of 150
Myr and for
three different stellar evolutionary models: $Z=0.008$ and $Z=0.019$,
both with modest convective overshoot, and $Z=0.019$ without
convective overshoot. Upper and lower errors are for 1$\sigma$
differences in $\tau_{\rm WD}$.

\section{Discussion\label{discussion}}
All eight of the observed WDs are much more massive than the typical
WD mass of $\approx 0.56\msun$ \citep{Bergeron1992}, and five of the
WDs have apparent distance moduli $(m-M)_V\approx 10.5$.  
It is therefore reasonable to
assume that at least five of these objects are members of NGC 2168.
LAWDS 11 is almost certainly \emph{not} a cluster member.  Its age is
likely older than that of the  cluster as a whole, and while the
uncertainties in the spectral fits leave open the possibility that it
is younger (which would require it to be hotter), the observed colors
are more consistent with the cooler (and older) interpretation.  The
distance modulus of this WD is foreground to the cluster by little
more than $1\sigma$, but the cooler temperature favored by the
\bv ~color again favors the foreground interpretation.

LAWDS
15 has a mass and age consistent with cluster membership, but the
calculated distance modulus is inconsistent with that of the other
WDs by $\sim 4\sigma_{M_V}$.  Assuming that the distance modulus is correct
and that there is no difference in reddening between LAWDS
15 and NGC 2168, LAWDS 15 is $\sim 185$ pc closer than the cluster.
Based on the spectral fits, LAWDS 15 has a cooling age of $\sim 50$
Myr.  If the WD has been moving away from the cluster since its
formation at a steady rate of only $0.4$ km/s, it will have covered
this distance.  Therefore, it is possible that LAWDS 15 was
once a cluster member and has escaped the cluster. 

The likelihood that a massive, hot WD
would be found foreground along the line of sight to the cluster
can be estimated from the luminosity function in Figure 16 of
\citet{Liebert2004}. The luminosity function gives the space density
of WDs with $M>0.8\msun$ and $\tau_{\rm cool}\leq 100$ Myr as $\sim
10^{-5.3} \,{\rm pc}^{-3}\,0.5\,{\rm mag}^{-1}$. This results
in an estimated 0.1 hot, massive WDs in the $\sim 30\arcmin\times
30\arcmin$ MOSAIC field to a distance of 1 kpc.  Therefore it is
unlikely, but not impossible, that LAWDS 15 is a field WD.   Based on
these arguments, we will consider LAWDS 15 
to be a cluster WD for this discussion.
For similar
arguments, we retain LAWDS 6 ($\sim 2\sigma_{M_V}$ closer than the
other cluster WDs) as a likely cluster member.

Figure~\ref{fig.ifm} shows the initial-final mass relation of these
seven cluster members, along with that of WDs from Hyades, Praesepe and
the Pleiades \citep{Claver2001} and from NGC 2516
\citep{Koester1996}.  Also shown are theoretical and semi-empirical
data from plots in \citet{Claver2001} and sources therein.  For the sake of
consistency, the initial and final masses of each WD from the
literature have been re-determined using our WD models and the
published \teff and \logg.  

From the figure, it can be seen that the NGC 2168 WDs form a monotonic
sequence of more massive WDs originating from more massive
progenitors.  This conclusion is robust, as changes in the
assumed age of NGC 2168 do not affect the relative positions of the
points, only the absolute initial masses. 
Figure~\ref{fig.ifm} also shows that the NGC 2168 initial-final
mass relation agrees with the empirical relation derived 
from previous clusters.
This need not be expected \emph{a priori}, as NGC 2168 has a
significantly lower metallicity than the other clusters in the
diagram.  Model core masses decrease with increasing metallicity, 
and the efficiency of mass loss processes could change with metallicity.  

LAWDS 15 and LAWDS 6, the potential escaped cluster WDs discussed above,
fit the observed initial-final mass relation,
providing additional evidence that they are cluster members.  If these
WDs are indeed escaped cluster members, they are crucial
pieces of evidence that WDs can receive velocity kicks during mass loss
and perhaps explain the observed deficit of WDs in other open clusters
\citep[see][and references therein]{Williams2004a}.  This and other
potential explanations for these objects (e.g. binarity) will be discussed more fully in
the later paper on WDs in NGC 2168.

\begin{figure}
\plotone{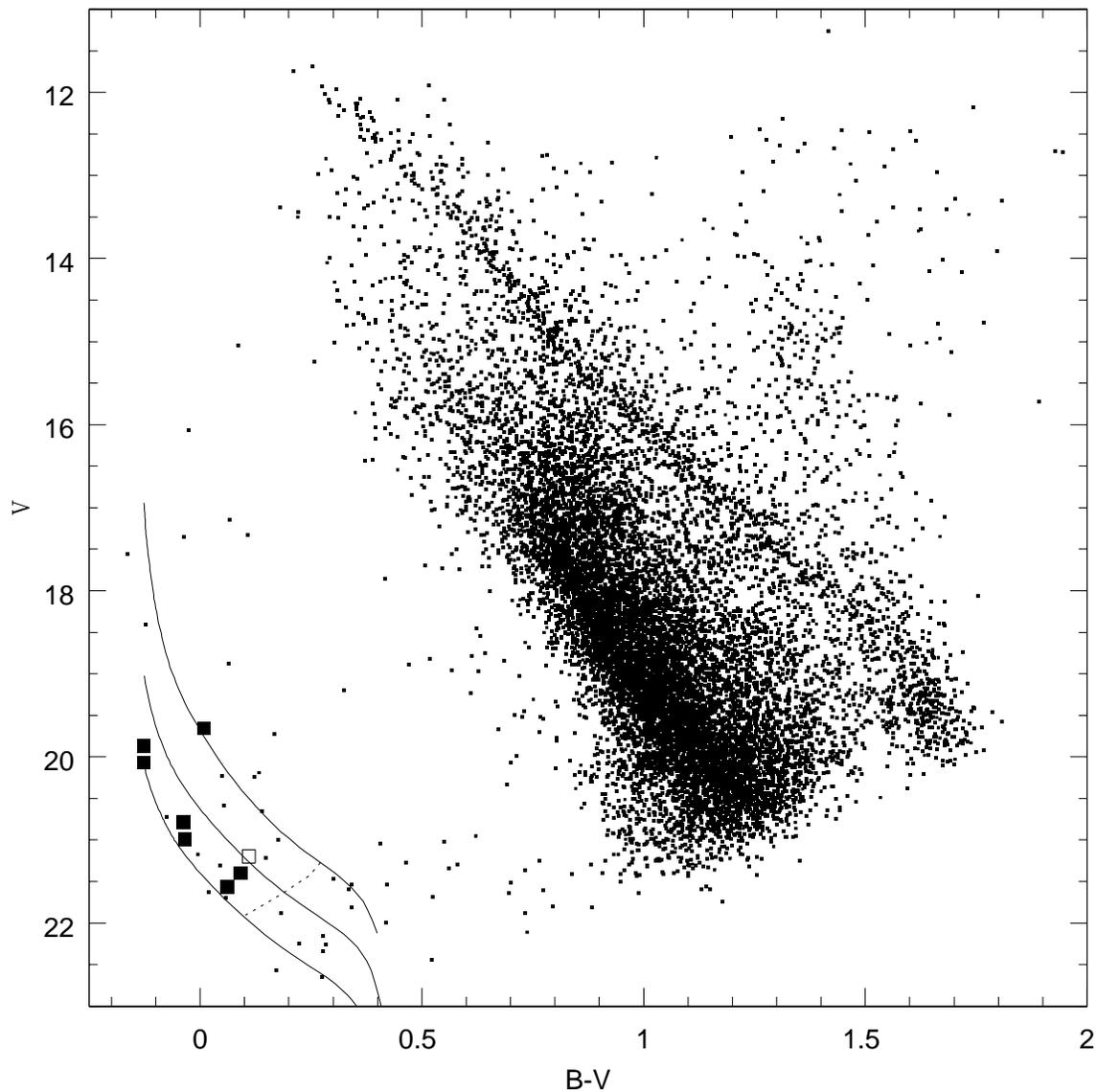}
\figcaption{$\bv,\ V$ color-magnitude diagram
for NGC 2168. Squares
  indicate WDs presented in this study, both cluster members (filled)
  and the non-member (open).  Solid curves indicate cooling curves for WDs
  with masses of 0.4\msun (top curve), 0.7\msun,
  and 1.0\msun(bottom curve) at the
  distance and reddening of NGC 2168.  The dotted line indicates the
  location of WDs with $\log \tau_{cool}=8.15$, the assumed age of NGC
  2168. \label{fig.cmd}} 
\end{figure}

\begin{figure}
\includegraphics[angle=270,width=6.25in]{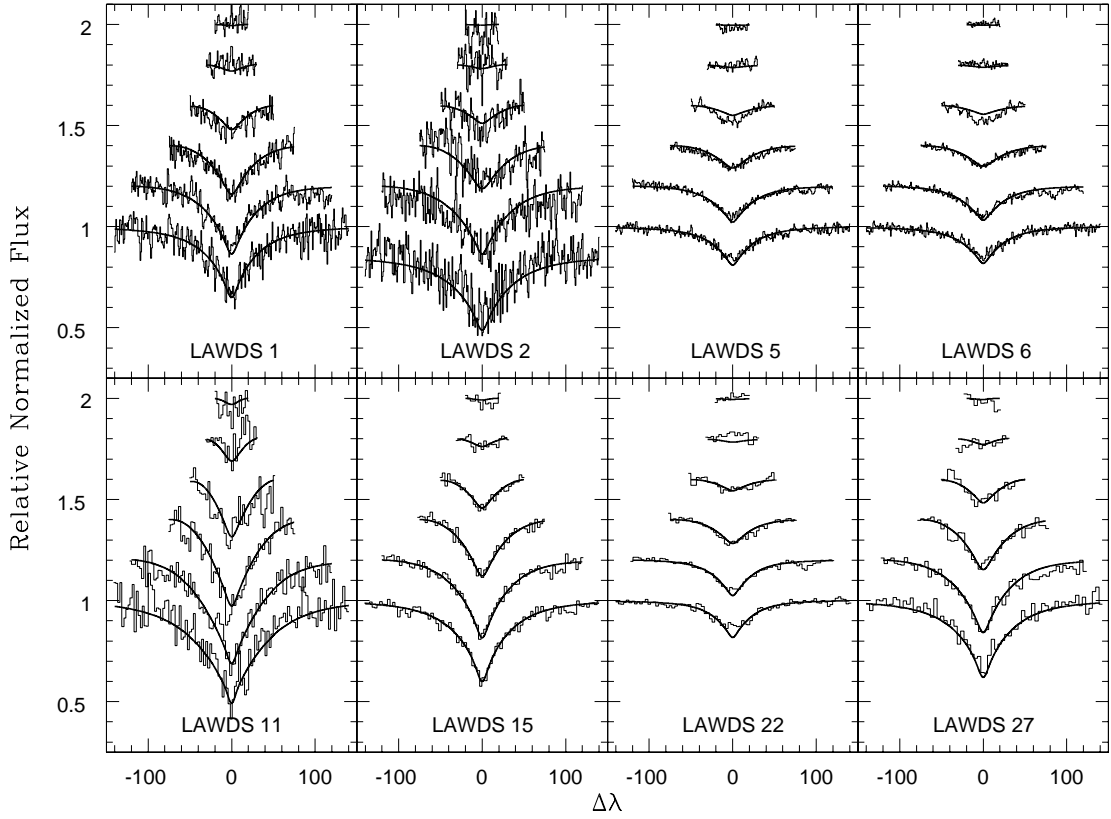}
\figcaption{Balmer line fits for WD candidates in NGC 2168. The figures
  show the Balmer lines from H$\beta$ (bottom) to H9 (top), with the
  curves showing the best-fit models. \label{fig.linefits}}
\end{figure}

\begin{figure}
\plotone{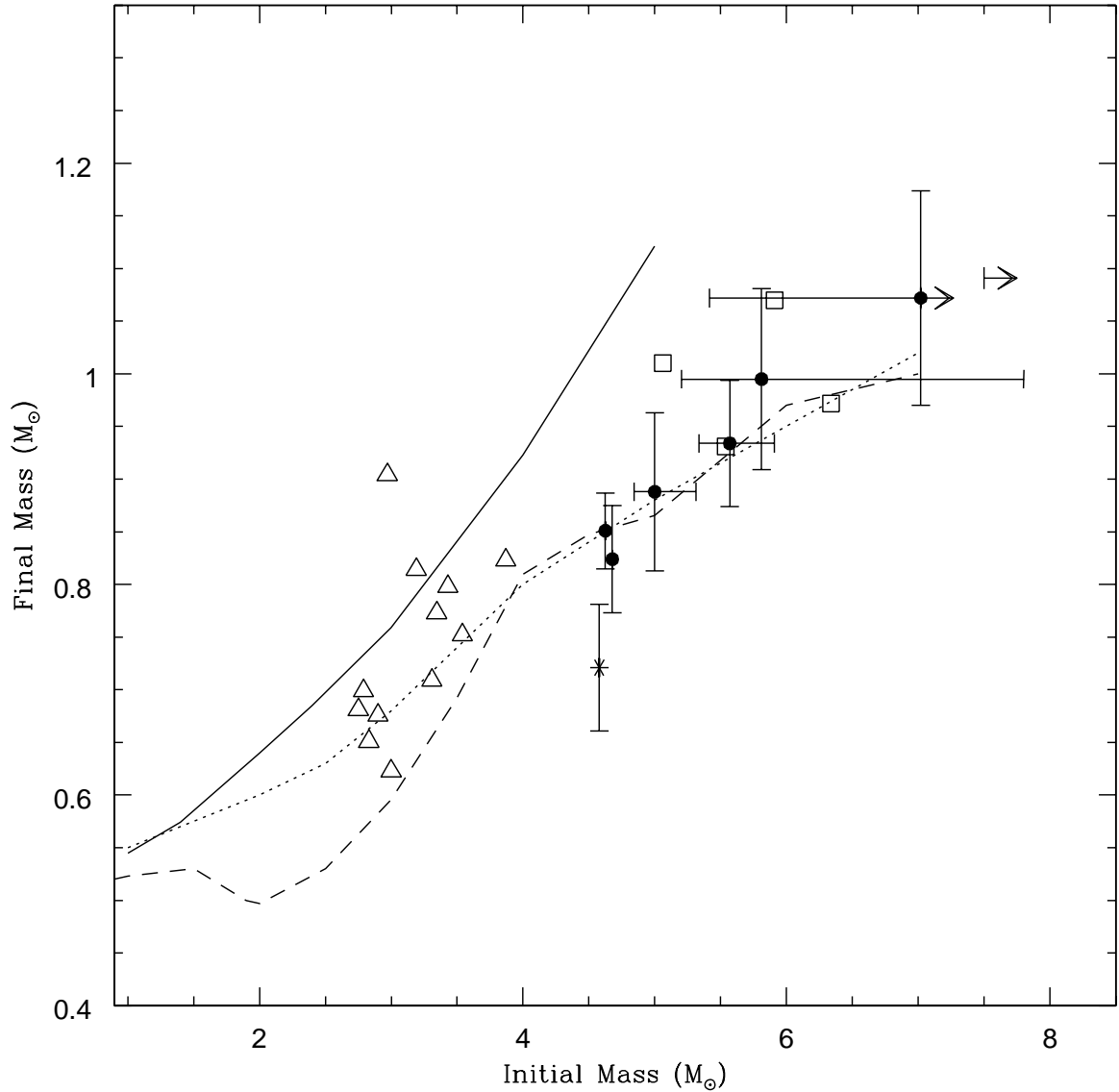}
\figcaption{Initial-final mass relation.  Points with error bars are WDs from
  this study; other points are from the Hyades and Praesepe (Claver et
  al.~2001, open triangles), NGC 2516 (Koester 1996, open squares) and
  the Pleiad WD (Claver et al.~2001, upper limit).  The starred point
  with error bars is LAWDS 22, discussed in the text. Curves represent
  the theoretical initial-final mass relation from \citet{Girardi2000}
  (solid), the core mass at the first thermal pulse from the same
  models (dashed), and the quasi-empirical relation from
  \citet{Weidemann2000} (dotted line). Slight horizontal offsets have been
  applied to the three WDs with $M_{\rm init}\approx 4.6\msun$ for
  the sake of clarity. Statistical error bars in the initial masses
  for these three WDs are smaller than the points.\label{fig.ifm}}
\end{figure}

The WD in Fig.~\ref{fig.ifm} with the apparently low final mass
is LAWDS 22.  As mentioned above and visible in Fig.~\ref{fig.linefits},
the spectral fitting did not converge satisfactorily, despite the
high signal-to-noise of the spectrum.  This star is a close visual
double with a redder companion ($V=19.12$, $\bv=1.27$, $\ub=0.93$)
2\arcsec to the north.
While resolved, this double is close enough that the spectrum of LAWDS
22 is likely contaminated by light from the neighbor star, resulting
in the 
unsatisfactory fit.  Light from the neighboring star may also be
contaminating the photometric colors of LAWDS 22, which would also explain
why the star lies redward of the 1\msun cooling track in Fig.~\ref{fig.cmd}.  
Given its measured \teff, LAWDS 22 likely suffers from an
extension of the systematic issue in the high-\teff WD 
spectral fits described above.  Other explanations for
the location of this point could include magnetic fields (although no
splitting is observed) or a low-mass, unresolved companion, but 
contamination by the neighboring star seems to be the most likely cause.

Based on the NGC 2168 data alone, it is possible to place lower limits
on the value of \mcrit.  Making the assumption that errors in the WD
ages in Table~\ref{tab.spec} are Gaussian, we calculate that the 
oldest cluster WD ages are $\log \tau_{\rm cool}\geq 7.76$
with 90\% confidence.  For $\log \tau_{\rm cl}=8.15$ and $Z=0.008$, this
corresponds to $\mcrit\geq 5.81\msun$ at a 90\% confidence level.
This value is in agreement with that obtained by \citet{Koester1996}.

Improved constraints on the initial-final mass relation, including its
intrinsic scatter and metallicity dependence, require improvements on
existing observations.  First and foremost, the ages of open clusters
such as NGC 2168 must be determined to higher precision.  Alternatives
to main-sequence fitting such as lithium depletion studies or
activity/rotation studies may provide the necessary constraints.
Second, large samples of WDs from individual open clusters are needed to
reduce the effect of systematics (such as errors in assumed ages) that
plague the comparison of inter-cluster samples.  The WD sample
presented here is a start to that end, and the initial-final mass
relation derived from these stars alone provides dramatic confirmation
of the existence of an initial-final mass relation, an idea that was
strongly hinted at from previous open cluster studies and from other
theoretical and observational work.  Assuming that the majority of the
remaining massive NGC 2168 WD candidates are cluster WDs,
planned spectroscopic observations of these objects will soon result
in a sample of nearly a dozen of WDs originating from a
single stellar population, permitting, for the first time, studies of
the intrinsic dispersion of the initial-final mass relation at high
masses.  

\acknowledgements
M.B.  and K.W. are grateful for support for this project in the
form of the National Science Foundation AST-0307492.
Any opinions, findings, and conclusions or recommendations expressed
in this material are those of the author(s) and do not necessarily
reflect the views of the National Science Foundation.

The authors would like to thank Matt Wood for freely providing
the WD evolutionary models and for many useful comments regarding this paper.
The authors also thank the anonymous referee for a well-considered
report that resulted in several improvements to this Letter.

Some of the data presented herein were obtained at the W.M. Keck
Observatory, which is operated as a scientific partnership among
the California Institute of Technology, the University of California
and the National Aeronautics and Space Administration. The Observatory
was made possible by the generous financial support of the W.M.
Keck Foundation.

The authors wish to recognize and acknowledge the very significant
cultural role and reverence that the summit of Mauna Kea has always
had within the indigenous Hawaiian community.  We are most fortunate
to have the opportunity to conduct observations from this mountain.

\clearpage

\begin{deluxetable}{llccccccc}
\rotate
\tabletypesize{\footnotesize}
\tablecolumns{9}
\tablecaption{WDs in the field of NGC 2168, photometry.\label{tab.phot}}
\tablewidth{0pt}
\tablehead{\colhead{LAWDS} & \colhead{McCook \& Sion} & \colhead{RA} & \colhead{Dec}& $V$ & $\sigma_V$ &  
  \colhead{$\bv$} & \colhead{$\sigma_{\bv}$} & \colhead{Previous} \\
  \colhead{ID} & \colhead{Designation\tablenotemark{a}} & \colhead{(J2000)} & \colhead{(J2000)} & & & & & \colhead{References}}
\startdata
\objectname{NGC 2168:LAWDS 1}  & \objectname{WD J0608+242}  & 6:08:38.79 & 24:15:06.9 & 20.989 & 0.019 & -0.035 & 0.028 &1\\
\objectname{NGC 2168:LAWDS 2}  & \objectname{WD J0608+241}  & 6:08:42.30 & 24:10:17.7 & 21.569 & 0.032 &  0.061 & 0.044 &\\
\objectname{NGC 2168:LAWDS 5}  & \objectname{WD J0609+244.1}& 6:09:11.54 & 24:27:20.9 & 20.065 & 0.017 & -0.128 & 0.024 &1,2\\
\objectname{NGC 2168:LAWDS 6}  & \objectname{WD J0609+244.2}& 6:09:23.48 & 24:27:22.0 & 19.863 & 0.016 & -0.128 & 0.023 &1,2\\
\objectname{NGC 2168:LAWDS 11} & \objectname{WD J0609+241}  & 6:09:42.79 & 24:11:05.4 & 21.198 & 0.025 &  0.110 & 0.037 &\\
\objectname{NGC 2168:LAWDS 15} & \objectname{WD J0609+240}  & 6:09:11.63 & 24:02:38.5 & 20.785 & 0.022 & -0.039 & 0.032 &\\
\objectname{NGC 2168:LAWDS 22} & \objectname{WD J0608+245}  & 6:08:24.65 & 24:33:47.6 & 19.657 & 0.016 &  0.008 & 0.023 &\\
\objectname{NGC 2168:LAWDS 27} & \objectname{WD J0609+243}  & 6:09:06.26 & 24:19:25.3 & 21.398 & 0.026 &  0.090 & 0.039 &3\\
\enddata
\tablenotetext{a}{Format as in \citet{McCook1999}}
\tablerefs{(1) \citet{Romanishin1980}, (2) \citet{Reimers1988}, (3) \citet{vonHippel2002}}
\end{deluxetable}

\clearpage

\begin{deluxetable}{lccccccccccc}
\rotate
\tabletypesize{\footnotesize}
\tablecolumns{12}
\tablecaption{WDs in the field of NGC 2168, spectral fits.\label{tab.spec}}
\tablewidth{0pt}
\tablehead{\colhead{ID} & \colhead{\teff} & \colhead{$d\teff$} & \colhead{\logg} & \colhead{$d\logg$} & 
   \colhead{$M_{\rm WD}$} & \colhead{$dM_{\rm WD}$} & \colhead{$\log(\tau_{\rm WD})$} & \colhead{$d\log(\tau_{\rm WD}$)} & 
   \colhead{$M_V$} & \colhead{$dM_V$} & \colhead{$(m-M)_V$}} 
\startdata
LAWDS 1  & 32400 &  512 & 8.40 & 0.125 & 0.888 & 0.075 & 7.386 & 0.208 & 10.454 & 0.215 & 10.535 \\
LAWDS 2  & 31700 & 1800 & 8.74 & 0.191 & 1.072 & 0.102 & 7.929 & 0.291 & 11.132 & 0.328 & 10.437 \\
LAWDS 5  & 52600 & 1160 & 8.24 & 0.095 & 0.824 & 0.051 & 6.158 & 0.025 &  9.542 & 0.184 & 10.523 \\
LAWDS 6  & 55200 &  897 & 8.28 & 0.065 & 0.851 & 0.036 & 6.094 & 0.027 &  9.564 & 0.108 & 10.299 \\
LAWDS 11 & 19900 & 2792 & 8.48 & 0.367 & 0.921 & 0.213 & 8.288 & 0.463 & 11.534 & 0.769 &  9.664 \\
LAWDS 15 & 29900 &  318 & 8.48 & 0.060 & 0.934 & 0.037 & 7.693 & 0.089 & 10.754 & 0.111 & 10.031 \\
LAWDS 22 & 54400 & 1203 & 8.04 & 0.121 & 0.721 & 0.060 & 6.169 & 0.054 &  9.155 & 0.219 & 10.502 \\
LAWDS 27 & 30900 &  500 & 8.58 & 0.164 & 0.995 & 0.086 & 7.760 & 0.220 & 10.866 & 0.310 & 10.532 \\
\enddata
\end{deluxetable}

\clearpage

\begin{deluxetable}{lccc}
\tablecolumns{4}
\tablecaption{Progenitor Masses for NGC 2168 WDs.\label{tab.masses}}
\tablewidth{0pt}
\tablehead{
    \colhead{ID} & \colhead{$Z=0.008$} & \colhead{$Z=0.019$} & \colhead{$Z=0.019$\tablenotemark{a}}\\
    & (\msun) &  (\msun) &  (\msun)}

\startdata
LAWDS 1  & $5.00^{+0.31}_{-0.16}$    & $5.08^{+0.29}_{-0.15}$    & $4.86^{+0.28}_{-0.14}$ \\ 
LAWDS 2  & $7.02^{+\infty}_{-1.60}$  & $6.96^{+\infty}_{-1.49}$  & $6.63^{+\infty}_{-1.40}$ \\
LAWDS 5  & $4.628^{+0.001}_{-0.001}$ & $4.734^{+0.002}_{-0.001}$ & $4.540^{+0.001}_{-0.001}$ \\
LAWDS 6  & $4.625^{+0.001}_{-0.001}$ & $4.732^{+0.001}_{-0.001}$ & $4.537^{+0.001}_{-0.010}$ \\
LAWDS 15 & $5.57^{+0.34}_{-0.24}$    & $5.62^{+0.31}_{-0.22}$    & $5.37^{+0.30}_{-0.21}$ \\
LAWDS 22 & $4.629^{+0.003}_{-0.003}$ & $4.735^{+0.003}_{-0.002}$ & $4.540^{+0.003}_{-0.002}$ \\
LAWDS 27 & $5.81^{+1.99}_{-0.61}$    & $5.84^{+1.81}_{-0.56}$    & $5.59^{+1.46}_{-0.54}$ \\
\enddata
\tablenotetext{a}{No convective overshoot}
\end{deluxetable}

\clearpage

\end{document}